\documentstyle[aps,prl,epsf,rotate]{revtex}
\begin{document}


\title{A Generalized Fokker-Planck Equation for the Ratchet 
Problem: Memory Effects}

\author{E. Ke{\c c}ecio{\u g}lu, B. Tanatar, and M. Cemal Yalab{\i}k}
\address{
Department of Physics, Bilkent University, Bilkent, 06533
Ankara, Turkey}
\maketitle
\begin{abstract}
We study stochastic ratchets with inertia in the limit where 
the time correlations become important. We have developed a 
Fokker-Planck type equation for the ratchet problem which 
includes the memory effects. It is tested with comparison 
to the Langevin simulations. The effect of the memory in 
the system is analyzed extensively. A positive feedback 
regime which manifests itself as instabilities is observed.
This may be regarded as an illustration of stochastic 
resonance.
\end{abstract}


The motion of a particle under the influence of random
forces and an asymmetric periodic potential has been attracting
a great deal of interest in recent years.\cite{1} The so-called
ratchet models exhibit a variety of interesting phenomena
stemming from nonequilibrium fluctuations. The most significant 
result of these investigations is the occurrence of a net 
macroscopic current\cite{2,3,4,5} which
depends on the spatial asymmetry of the external potential 
as well as the statistical properties of the fluctuating forces. 
The stochastic ratchets are important in understanding
biological systems such as molecular motors\cite{2,3,6} and they 
also find application for electrons in superlattices, quantum 
dots, Josephson junctions, and atomic systems.\cite{7,8} Recent 
experiments\cite{9} making use of nanolitography techniques 
provide a wealth of possibilities in the study of quantum 
properties.\cite{10}

In this work we study the effects of memory friction for a
Brownian particle subject to spatial and temporal
forces. When the dissipative effects of the fluctuating (random)
force are allowed to depend on the system's past behavior, the 
memory function replaces the constant friction term and a
generalized Langevin equation results.\cite{11} Previous studies of
thermal ratchets making use of the memory function or correlated
noise relied on the simulation of Langevin equation to obtain 
transport properties.\cite{12,13,14,15,16}
It has been remarked that a corresponding Fokker-Planck 
equation\cite{17} (FPE) describing the dynamics of the system 
is difficult to construct owing to the non-Markovian nature 
of the underlying process. There has been several attempts to 
obtain a FPE with colored Gaussian noise.\cite{18}
Fokker-Planck equation description in this context
presents the global characteristics and in particular the 
current density which reveals the most interesting
properties of thermal ratchets such as rectification and 
current reversal.\cite{5,19} Our chief aim here is to obtain a
FPE in the presence of memory effects in an approximate way. 
Introducing some simplifying assumptions we
form an integro-differential equation satisfied by the
probability density which may be construed as a
generalized Fokker-Planck equation. We remark that the 
non-Markovian nature of the problem is preserved 
in the final FPE we obtain. It has been known 
that for a non-Markovian system, the is no generic way of
obtaining a FPE if the system does not exhibit steady state 
solutions. In our case, we know from the outset that steady 
state solutions exist through Langevin simulations, even in the
non-Markovian limit (i.e. with memory effects).
Thus, deriving a FPE equation for this 
system does not contradict the well known features of the 
non-Markovian problems.\cite{17} 
We demonstrate by numerical calculations that the same
qualitative and quantitative results can be obtained as those
from the generalized Langevin equation. 

Our application of the generalized FPE to the ratchet problem
shows interesting properties resulting from the memory effects. 
In particular, we find a regime in the current characteristics 
exhibiting instabilities. 
In the following we first sketch a derivation of the generalized FPE
in the form of an integro-differential equation which has
general applicability to the ratchet problem. We then present
and discuss our results emphasizing the memory effects. 

Our aim is to construct a Fokker-Planck equation for stochastic
ratchets including the memory effects.
We begin with the basic equation of motion governing the 
dynamics of a Brownian particle, given by the generalized Langevin 
equation
\begin{equation}
\frac{dv}{dt} =-\int_{-\infty}^{t} \mu(t-t')\,v(t')\,dt'+
f(t)-\frac{dV(x)}{dx}+\xi(t)\, ,
\end{equation}
where $x(t)$ and $v(t)=dx/dt$, respectively, are the position 
and velocity of the particle, $V(x)$ is the asymmetric periodic
potential, $f(t)$ is a time-dependent external driving force,
and $\xi(t)$ is the stochastic force with zero mean. Each term
in Eq.\,(1) is scaled with the particle mass $m$. We note
that the above description is quite general and encompasses a 
variety of situations already covered in previous 
studies.\cite{2,3,4,5,7} Assuming we know the solution at 
some time $t$, it is possible to construct the solution at 
time $t+dt$. Let the solution for the position (velocity) be 
$x$\,($v$) and $x'$\,($v'$) at time $t$ and $t+dt$, respectively. 
Then it is straightforward to construct the following set of 
equations:
\begin{equation}
v'=v- \left[\int_{-\infty}^{t} \mu (t-t')~v(t')~dt'-
f(t)+\frac{dV(x)}{dx}\right]\,\Delta t + \xi(t)\,(\Delta t)^{1/2}\,
,\qquad\hbox{and}\qquad x'=x+v\Delta t\, .
\end{equation}
The power $1/2$ of $\Delta t$ multiplying the random force is 
a requirement imposed by the zero-mean property of $\xi(t)$, 
so that the lowest order contribution of the noise term is in 
second order.
                     
Our goal is to construct an equation for the probability 
density function $P(v,x,t)$. Since we can express $v'$ and $x'$
at time $t+dt$ when we know $v$ and $x$ at time $t$, it should 
also be possible to find a solution for $P(v',x',t+dt)$ when 
we know $P(v,x,t)$. In other words, 
by using the fact that the total probability is conserved, we
aim at finding how $P(v,x,t)$ transforms to $P(v',x',t+dt)$.
To this end, we first discretize the integral in Eq.\,(2), 
so that 
\begin{equation}
-\int_{-\infty}^{t} \mu (t-t')~v(t')~dt'=
-\sum_{n=0}^{\infty}\mu(n\Delta t')\,v(t-n\Delta t') \Delta t'\,
,
\end{equation}
with $\mu(0)\Delta t'$ approaching $\gamma$ as  
$\Delta t'\rightarrow 0$. Here $\gamma$ is the friction 
constant in the absence of memory correlations in the
dissipation term.
In order to simplify the notation we define $v_n$ as 
$v(t-n\Delta t')$, $v$ as velocity at time $t$, and $v'$ 
as velocity at time $t+\Delta t$. At this point, remembering
that all $v_n$'s, $v$, $v'$, $x$, and $x'$ are random 
variables, we write
\begin{equation}
P(v',x',t+\Delta t)=\int dx \int dv \int d\xi  \int 
\left[\prod_n dv_n\right]\,\left[\prod_n P_n(v_n)\right]\,P(v,x,t)
\,P_\xi\,\delta(x'-x-v~\Delta t ) 
\end{equation}
\begin{displaymath}
\times~\delta\left(v'-v+\left[\sum_{n=0}^{\infty}\mu(n\Delta t')v_n 
\Delta t'-F(x)-f(t)\right]\Delta t - \xi(t) (\Delta t)^{1/2}\right)\,
.
\end{displaymath}
Here $P_n$, is the probability density function for the random 
variable $v_n$ and similarly $P_\xi$ is the probability density 
function for $\xi$. The product over $P_n(v_n)$ indicates an 
assumption of statistical independence of the velocities ${v_n}$ 
at different times, although such velocities are clearly 
expected to be correlated. Nevertheless, this approximation is 
expected to be valid for velocity distributions near steady state, 
and enables us to proceed with the algebra.
Performing the integrations over $v$ and $x$, keeping only terms
of order $\Delta t$, and considering the limits
$\Delta t'\rightarrow 0$ and $\Delta t\rightarrow 0$, we
find
\begin{equation}
\frac{\partial P}{\partial t} = \gamma P(v,x,t)  +  
\frac{\partial P}{\partial v} \left[\int_{-\infty}^{t} \mu(t-t')
  \bar{v}(t')~dt'-F(x)-f(t)+ \gamma (v-\bar{v})\right]- 
\frac{\partial P}{\partial x} v +
D\,\frac{\partial^2P}{\partial v^2}\, .
\label{FP}
\end{equation}
Here $\bar{v}(t)$ is the average velocity at a
given time, $\gamma$ is equal to $\mu(0) \Delta t'$ as 
$\Delta t'\rightarrow 0$ as defined earlier, and $D$ is the
diffusion constant. 
Equation\,(5) is the main result of this paper. Under the
assumptions set out in the preceding paragraphs (i.e.
statistical independence), it describes
the dynamics of stochastic ratchets when the memory effects are
included. We recover the conventional FPE if the 
memory function is chosen to be a delta function. In the
following we demonstrate that the solutions of Eq.\,(5) 
are in good agreement with the Langevin simulation results. 

We have solved the Fokker-Planck and Langevin equations 
numerically to test how well our proposed FPE 
describes the phenomenology of thermal ratchets. We have found
that Eq.\,(5) reproduces all the established results
such as current reversal and noise rectification rather well. 
In the results to be discussed below, we have specifically used
$f(t) = A\sin{(\omega t)}$ with $A=1.0$ and 
$\omega=0.2$ for the driving force. The ratchet potential is 
taken to be $V(x)=b_0 \sin{(2\pi x /L)}+b_1 \sin{(4\pi x /L)}$,
with $b_0$, $b_1$, and $L$ are constants as has been used 
by others.\cite{4,5,12} We model 
the memory function in the form $\mu(t)=(\gamma/\tau)
\exp{(-|t|/\tau)}$, where $\tau$ is the correlation time. 
Finally, $\xi(t)$ is a Gaussian random variable whose average 
is zero and time correlation is given by the fluctuation-dissipation 
theorem $\langle\xi(t)\xi(t')\rangle=2D\mu(t-t')$, for
$t>t'$, and approaches $\langle\xi(t)\xi(t')\rangle =2\gamma D 
\delta (t-t')$ as $\tau$ approaches zero.

We first demonstrate that our proposed Fokker-Planck equation 
[Eq.\,(5)] and Langevin equation [Eq.\,(1)] yield 
the same current value for the ratchet problem. As seen 
from these two equations, we can construct a solution for 
both of them if we know the solution at some initial time. 
We have solved both equations using a simple finite element 
algorithm. In Fig.\,1 we illustrate our results for the 
time dependence of the velocity. 
Figure 1a shows the instantaneous velocities in both approaches
and we observe that the solutions are similar. 
The effective or average velocity of the particle 
corresponding to long time behavior is shown in Fig.\,1b.
A simple running average algorithm is used to 
find the effective (average) velocities. In fact, when 
properly averaged over the periods of oscillations, the 
long-time behaviors are just straight lines. Nevertheless, 
it is seen that the effective 
velocity, or equivalently the current  which is the physical 
observable, is the same for both methods. It is evident in
Fig.\,1 that the FPE results, approach the Langevin 
equation values with increasing time.

We now turn to the effects of 
memory correlation in the ratchet problem. In this regard,
we have analyzed the ratchet problem defined in 
the previous section for various values of the correlation 
time $\tau$ and strength of the correlations $\gamma$.
In Fig.\,2 we depict $x(t)$ for various values 
of $\tau$. It is evident that as 
we increase $\tau$, because of negative feedback of the 
correlation to the system, the current is decreased (current is 
proportional to the slope of the curves). When we explore
the large $\tau$ regime such that it becomes comparable to 
the period of oscillations in $x(t)$, we find a different
behavior. In Fig.\,3 we again show $x(t)$ for different values 
of $\tau$. The topmost 
curve ($\tau=0$) is enlarged many times in order to show the 
differences between the case when $\tau=0$ and when $\tau$ is 
large. It is observed that as $\tau$ becomes comparable 
to the period of oscillations of the $\tau=0$ case, there 
appears some instabilities in the system stemming from the
positive feedback of the memory. The memory term in 
the definition of the friction now behaves as if it is not a 
friction term but rather as a driving term in resonance with the 
actual frequency of the oscillations. Stochastic resonances in
ratchet problems or in a class of Fokker-Planck equations
have been reported.\cite{20} In the absence of a time-dependent
driving force, no instabilities are observed, consistent with
the known results.\cite{21} We have checked that similar 
instabilities are also obtained by direct Langevin simulations.
A further question is how the system behaves as the strength 
of the total friction changes. In Fig.\,4 the time development
of the position $x(t)$ for various values of $\gamma$ when
an instability is encountered ($\tau=30$) is shown. 
We observe that as the magnitude of $\gamma$ increases, 
the period of oscillations decreases. In the inset of Fig.\,4 
we show the period of oscillations as a function of $\gamma$
which indicates an exponential decrease. This again illustrates 
the positive feedback in our ratchet system.

In summary, we have developed a Fokker-Planck equation
corresponding to a generalized Langevin equation
to describe the dynamics of a Brownian particle under the
influence of external potentials and memory effects. The
proposed FPE reproduces the established results and reduces
to the known case in the white-noise limit when the memory 
effect is absent. We have demonstrated the adequacy of 
generalized FPE description by comparing the solutions to 
that of the Langevin equation simulations. The correlated 
fluctuating force gives rise to resonances for certain choice 
of the parameters. Finally, since Eq.\,(1) can also be 
interpreted\cite{5,21,22} as the Heisenberg equation for 
a quantum particle coupled to a heat bath, our corresponding 
FPE may be explored to study the quantum ratchets.

This work was partially supported by the
Scientific and Technical Research Council of Turkey (TUBITAK)
under Grant No. TBAG-1662. We thank {\"O}. T{\"u}rel for
discussions.

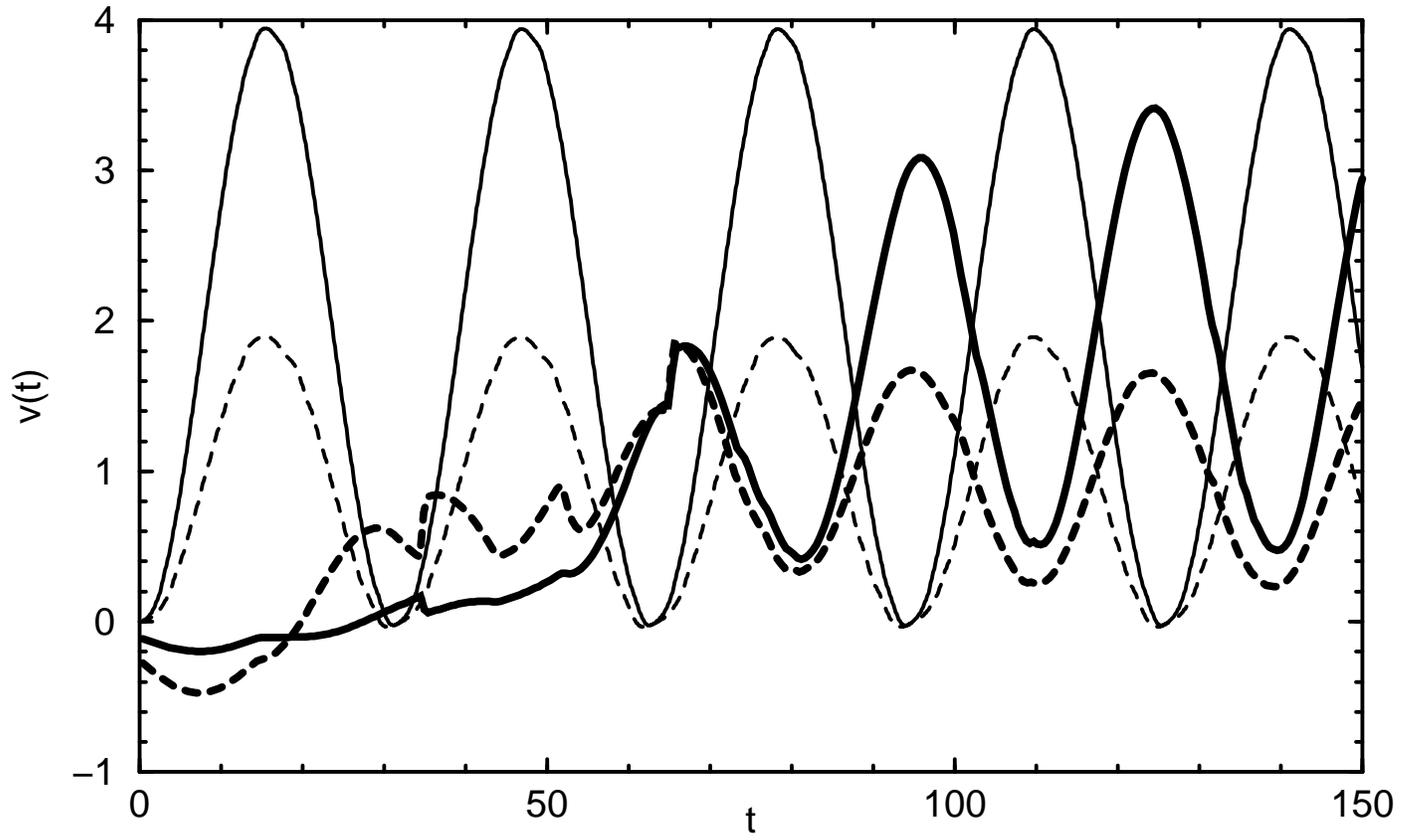
\begin{figure}[h]
\caption{(a) The instantaneous velocity calculated from the Langevin 
simulation (thin lines) and from the solution of generalized
Fokker-Planck equation (thick lines). The solid and dashed lines
indicate the amplitude $A=1$ and $A=0.5$, respectively, of the 
driving force f(t). (b) The 
corresponding average velocities for the same parameters.}
\end{figure}

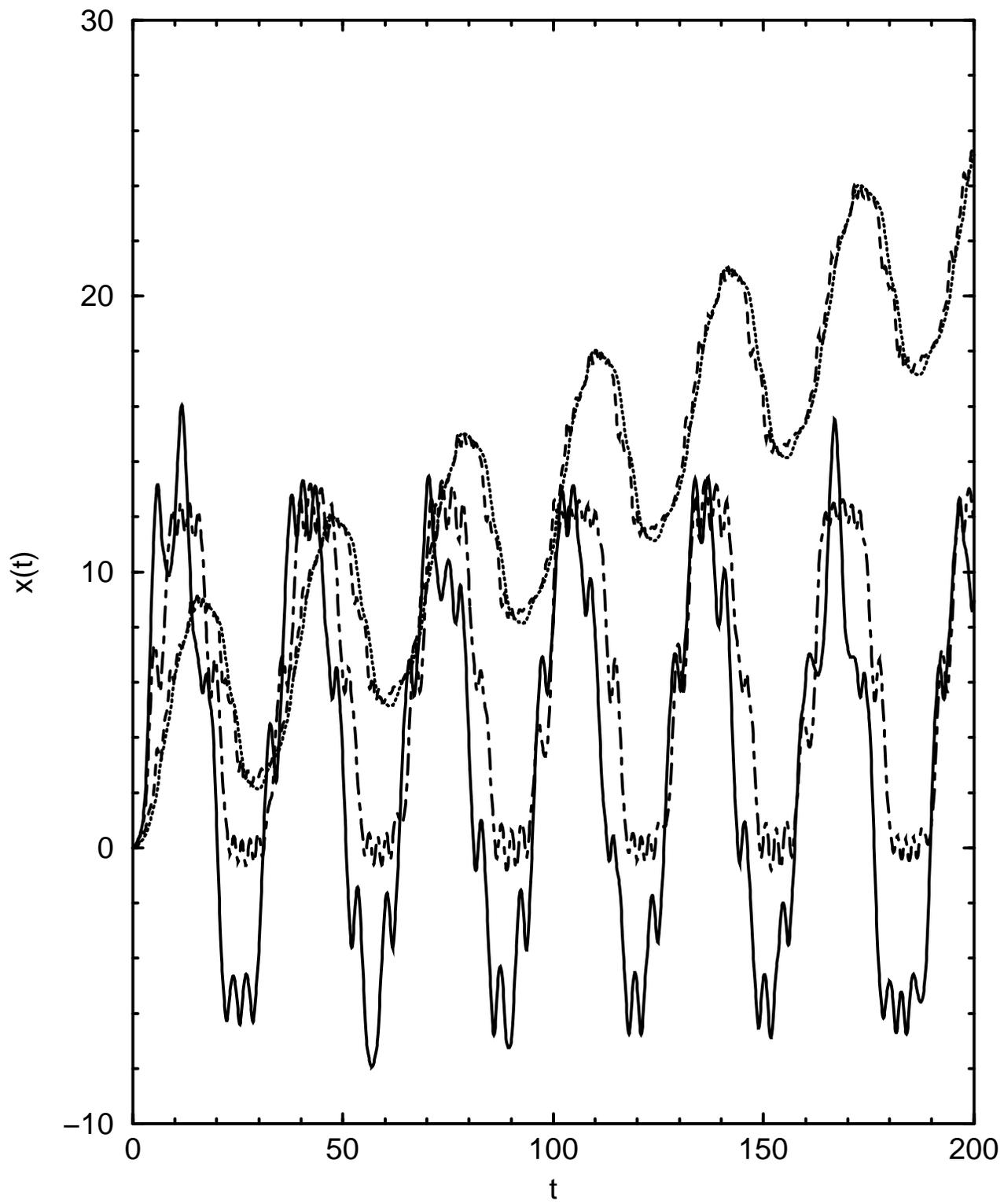
\begin{figure}[h]
\caption{The position variable $x(t)$ for various values of
the correlation time $\tau$ at fixed $\gamma=1$. The dotted,
dashed, dot-dashed, and solid lines indicate $\tau=0$, 1, 
5, and 10, respectively.}
\end{figure}

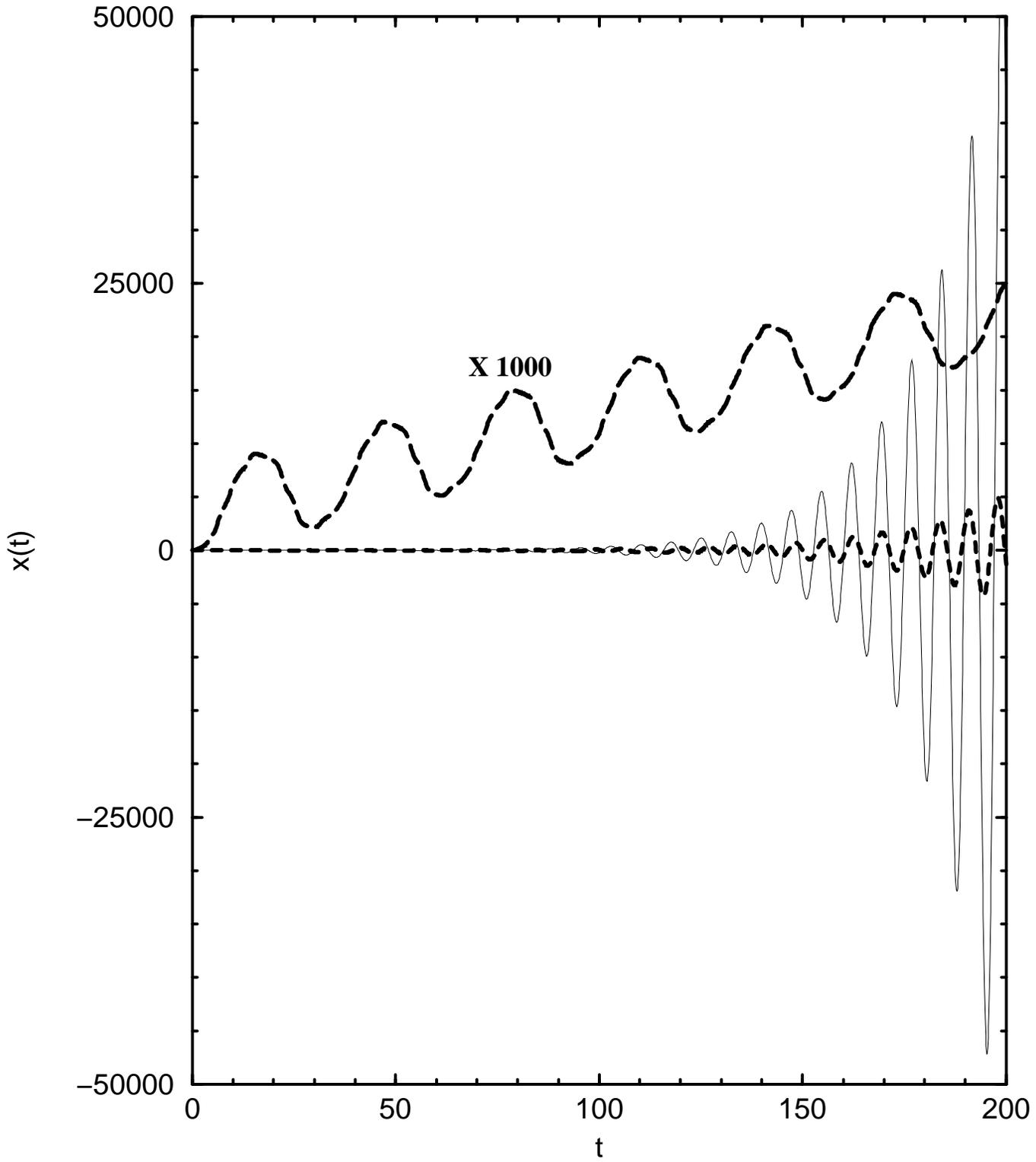
\begin{figure}[h]
\caption{The position variable $x(t)$ for various values of 
the correlation time $\tau$ at $\gamma=1$. The dotted and solid
lines indicate $\tau=18$ and $\tau=30$, respectively. $\tau=0$
(dashed line) case is also shown for reference.}
\end{figure}

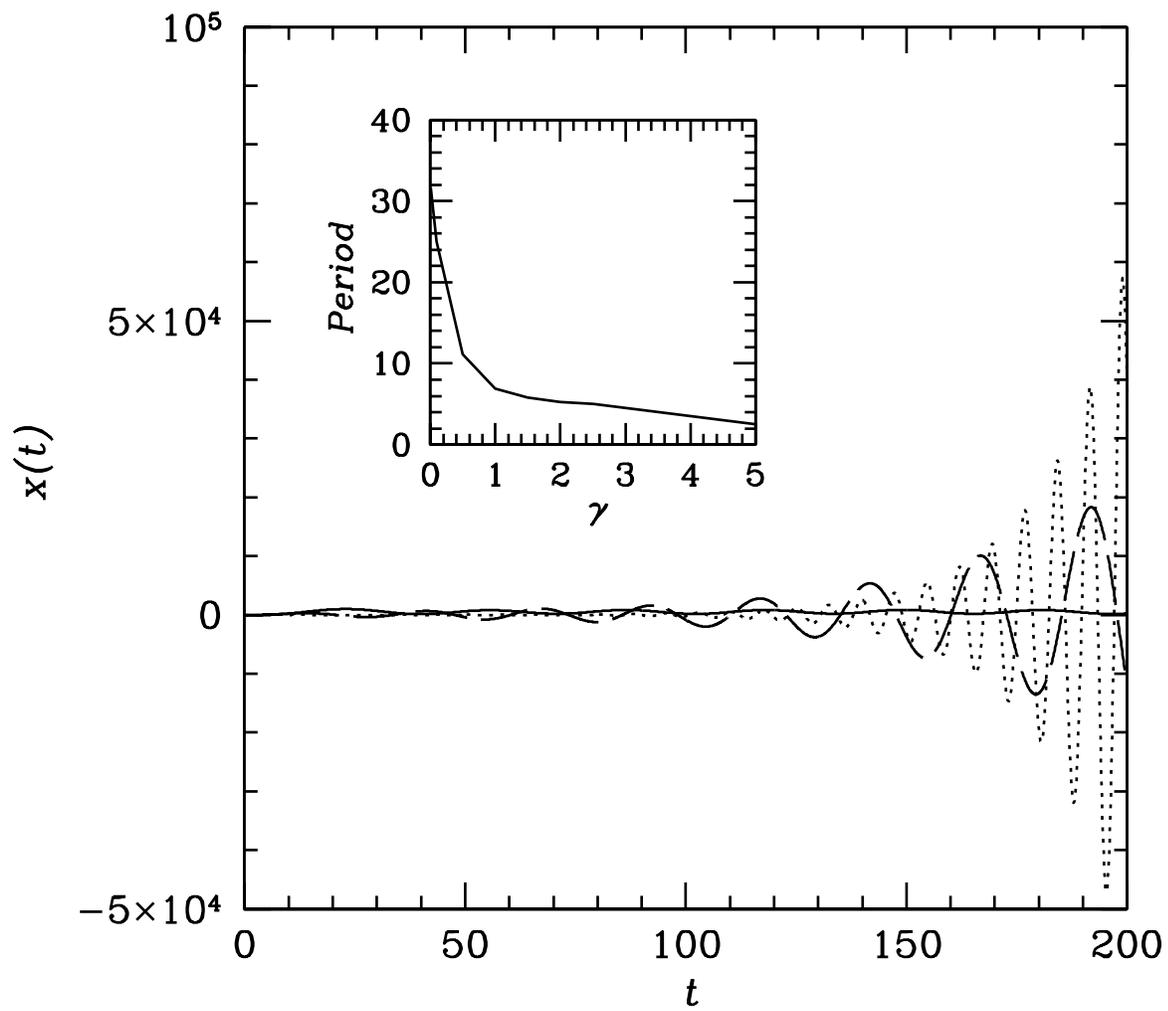
\begin{figure}[h]
\caption{The position variable $x(t)$ at $\tau=30$. The solid, 
dashed and dotted lines indicate $\gamma=0.01$,  $\gamma=0.1$, and  
$\gamma=1.0$, respectively.
The inset shows the dependence of the oscillation period on
$\gamma$.}
\end{figure}

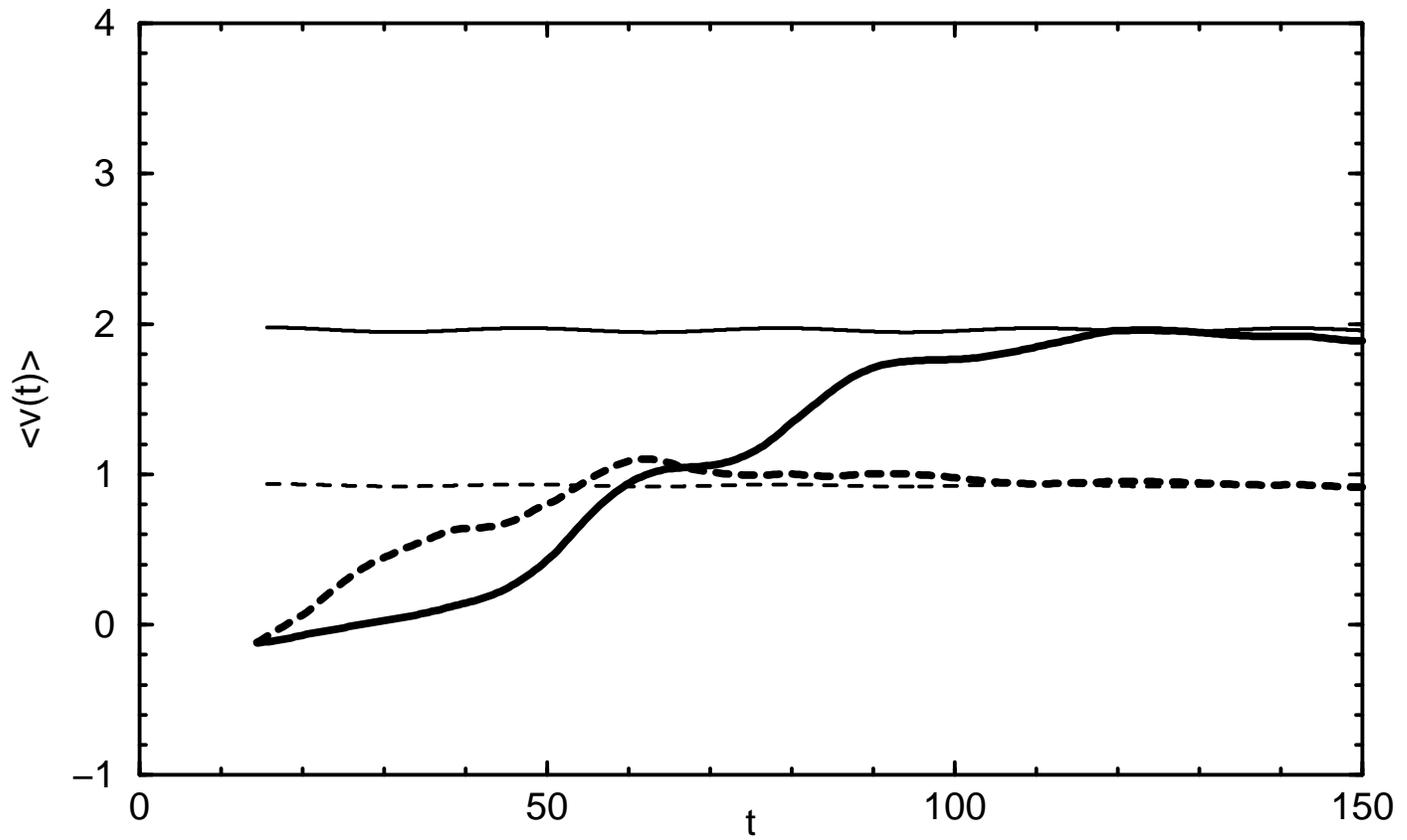
\end{document}